\input phyzzx
%\input epsf.tex
%\epsfverbosetrue
\hfuzz 50pt

%%%%%%%
\font\mybb=msbm10 at 12pt
\def\bb#1{\hbox{\mybb#1}}
\def\bZ {\bb{Z}}
\def\bC {\bb{C}}
\def\bR {\bb{R}}
\def\bE {\bb{E}}

\def\bH {\bb{H}}

\def\bfomeg{\omega\kern-7.0pt\omega}
\def\bfOmeg{\Omega\kern-8.0pt\Omega}

\def\os{{\buildrel  o\over s}}
\def\of{{\buildrel  o\over f}}
\def\oo{{\buildrel  o\over \omega}}
%%%%%%%%%%%%

%%%%%%%%%%%%%%%%%%%%%%%%%%%%%%%%%%%%%%%%%%%%%%%%%%%%%%%%%%%%%%%%%%%%%%%%%%%%%
\REF\gppkt{G. Papadopoulos \& P.K. Townsend,  
{\sl Intersecting M-branes}, Phys. Lett.
 {\bf B380} (1996) 273.}
\REF\revone{D. Youm, {\sl Black Holes and Solitons 
in String Theory}; hep-th/9710046.}
\REF\witten{E.Witten {\sl Solutions of four-dimensional 
field theories via M-theory}, Nucl. Phys.
{\bf B500} (1997) 3.}
\REF\tseytlinb{A.A. Tseytlin, {\sl Harmonic 
Superposition of M-Branes}, Nucl. Phys.{\bf B475}
(1996) 149.}
\REF\jerome{J.P. Gauntlett, D.A. Kastor \& J. 
Traschen, {\sl Overlapping Branes in M-theory}
 Nucl. Phys.  {\bf B478} (1996) 544.}
\REF\gpgpb{G. Papadopoulos \& P.K. Townsend, 
{\sl Kaluza-Klein on the brane}, Phys. Lett.
{\bf B393} (1997) 59.}
\REF\eric{E. Bergshoeff, M. de Roo, E. Eyras, B. Janssen 
and J.P. van de Schaar,
{\sl Multiple Intersections of D-branes and M-branes}, 
Nucl. Phys. {\bf B494} (1997) 119;
hep-th/9612095.}
\REF\leigh{M. Berkooz, M.R. Douglas \& R.G. Leigh, 
{\it Branes Intersecting at
Angles}, Nucl. Phys.  {\bf B480} (1996) 265.}
\REF\ohtatown{N. Ohta \& P.K. Townsend, {\sl Supersymmetry of M-branes
at Angles}, Phys. Lett. {\bf B418} (1998) 77; hep-th/9710129.}
\REF\ptg{J. P. Gauntlett, G.W. Gibbons, 
G. Papadopoulos \& P.K. Townsend, {\sl
Hyper-K\"ahler Manifolds and Multiply 
Intersecting Branes}, Nucl. Phys. {\bf B500} (1997)
133.}
\REF\ptb{G. Papadopoulos \& A. Teschendorff, 
{\it Multi-angle five-brane intersections},
Phys. Lett. in press; hep-th/9806191.}
\REF\cvetic { K. Behrndt \& M. Cveti\v c, 
{\sl BPS-Saturated Bound States of Tilted
p-Branes in Type II String Theory}, Phys. Rev. 
{\bf D56} (1997) 1188.}
\REF\myers{J.C. Breckenridge, G. Michaud and
 R.C. Myers, {\sl New Angles on Branes}, Phys. Rev. 
{\bf D56} (1997) 5172;
hep-th/9703041.}
\REF\costa{ M. Costa \& M. Cveti\v c, 
{\sl Non-Threshold D-Brane Bound States and
Black Holes with Non-Zero Entropy},  Phys. Rev. {\bf D56} 
(1997) 4834; hep-th/9703204.}
\REF\larsen{V. Balasubramanian, F. Larsen \& R.G. Leigh, 
{\sl Branes at Angles and
Black Holes}, Phys. Rev. {\bf D57} (1998) 3509;  hep-th/9704143.}
\REF\myersb{G. Michaud and R.C. Myers, {\sl Hermitian 
D-Brane Solutions}, Phys. Rev. {\bf D56} (1997) 3698;
hep-th/9705079.}
\REF\papasa{E. Bergshoeff, R. Kallosh, T. Ortin \& 
G. Papadopoulos, {\sl Kappa-symmetry,
supersymmetry and intersecting branes}, Nucl.
Phys. {\bf B502} (1997) 149; hep-th/9705040.}
\REF\papas{G.W. Gibbons \& G. Papadopoulos, {\it Calibrations
 and Intersecting Branes}, Commun.
Math. Phys. to appear, hep-th/9803163.}
\REF\west{J.P. Gauntlett, N.D. Lambert \& P.C. West, 
{\it Branes and Calibrated
Geometries}, hep-th/9803216.}
\REF\spence{B.S. Acharya, J.M. Figueroa-O' Farrill 
\& B. Spence, {\sl Planes,
Branes and Automorphisms I. Static Branes}, 
JHEP {\bf 07} (1998) 004; hep-th/9805073.}
\REF\figue{ J.M. Figueroa-O' Farrill, {\sl Intersecting Brane
Geometries}, hep-th/9806040.}
\REF\calia {R. Harvey and H. Blaine Lawson, Jr,
{\sl Calibrated Geometries}, Acta Math. {\bf
148} (1982) 47.}
\REF\calib{F.R. Harvey, {\sl Spinors and Calibrations}, 
Academic Press (1990), New York.}
\REF\calic{J. Darok, R. Harvey \& F. Morgan, {\it Calibrations on $\bR^8$},
Tranc. Am. Math. Soc., Vol. 307 {\bf 1} (1988), 1.}
\REF\donaldson{S.K. Donaldson \& R.P. Thomas, {\sl Gauge Theory in Higher
Dimensions}, Oxford University preprint.}
\REF\gpt{G. Papadopoulos \& A. Teschendorff, 
{\it Instantons at Angles},
Phys. Lett. {\bf B419} (1998) 115; hep-th/9708116.}
\REF\kent{E. Corrigan, P. Goddard \& A. Kent, {\sl Some Comments 
on the ADHM Construction
in 4k Dimensions}, Commun. Math. Phys. {\bf 100} (1985) 1.}
\REF\kelly {G.W. Gibbons, G. Papadopoulos \& K.S. Stelle, 
{\sl HKT and OKT Geometries
on Black Hole Soliton Moduli Spaces}, Nucl. 
Phys. {\bf B508} (1997) 623;
hep-th/9706207.}
\REF\yi{K. Lee, E. Weinberg \& P. Yi, {\sl The Moduli 
Space of Many BPS Monopoles
for Arbitrary Gauge Groups}, Phys. Rev. 
{\bf D54} (1996) 1633; hep-th/9602167.}
\REF\gwgnm{G.W. Gibbons \& N.S. Manton, {\sl The 
Moduli Metric for Well Separated BPS
Monopoles}, Phys. Lett. {\bf B356} (1995) 32.}
\REF\howec{P.S. Howe \& G. Papadopoulos, 
{ \sl Twistor Spaces for HKT manifolds},
Phys. Lett.  {\bf B379} (1996) 80.}
\REF\strominger{A. Strominger, {\sl Superstrings with Torsion}, Nucl.
Phys. {\bf B274} (1986) 253.}
\REF\callan{C. G. Callan,Jr, J.A. Harvey, A. Strominger,
{\sl Supersymmetric String Solitons}, hep-th/9112030; {\sl World
Sheet Approach to Heterotic Instantons and Solitons}, Nucl. Phys. 
{\bf B359} (1991) 611.}
\REF\hull{S.J. Gates, C.M. Hull \& M. Ro\v cek, 
{\sl Twisted Multiplets and New
Supersymmetric Nonlinear Sigma Models}, Nucl. Phys. 
{\bf B248} (1984) 157.}
\REF\howea {P.S. Howe \& G. Papadopoulos, {\sl Ultraviolet
 Behaviour of Two-Dimensional
Nonlinear Sigma Models}, Nucl. Phys. {\bf B289} (1987) 264; 
{\sl Further Remarks on the
Geometry of Two-Dimensional Nonlinear Sigma Models}, Class. 
Quantum Grav. {\bf 5} (1988)
1647.}
\REF\howeb{P.S. Howe \& G. Papadopoulos, {\sl Finiteness 
and Anomalies in (4,0)
Supersymmetric Sigma Models},  Nucl. Phys. 
{\bf B381} (1992) 360. }
\REF\hullwitten{C.M. Hull  \& E. Witten, {\sl Supersymmetric Sigma
Models and the Heterotic String},
Phys. Lett. {\bf B160} (1985) 398.}
\REF\kach {S. Kachru \& E. Silverstein, {\it 4d Conformal Field Theories
 and Strings on
Orbifolds}, hep-th/9802183.}

%%%%%%%%%%%%%%%%%%%%%%%%%%%%%%%%%

%%%%%%%%%%%%%%%%%%%%%%%%%%%%%%%%%%%%%%%%%%%%%%%%%%%%%%%%%%%%%%%%%%%%

\Pubnum{ \vbox{ \hbox{DAMTP-R/98/148} } }
\date{2 November 1998}
\pubtype{}

\titlepage

\title {\bf Grassmannians, Calibrations and Five-Brane Intersections}

\author{G. PAPADOPOULOS and A. TESCHENDORFF}
\address{DAMTP,\break University of Cambridge,
\break
Silver Street,
\break
Cambridge CB3 9EW, U.K.}
%\andauthor{}
%\address{}

\abstract{We present a geometric construction of a new
class of hyper-K\"ahler manifolds with torsion. This involves
the superposition of the four-dimensional hyper-K\"ahler
geometry with torsion associated with the NS-5-brane
 along quaternionic planes in
$\bH^k$.  We find  the moduli space of these geometries and show
that it can be constructed using the bundle space of the
canonical quaternionic line bundle over a quaternionic projective space.
We also investigate several special cases which are associated
with certain classes of quaternionic planes in $\bH^k$. We then 
show that the eight-dimensional geometries we have found can be
 constructed using quaternionic
calibrations. We  generalize  our construction to 
superpose the same  four-dimensional
hyper-K\"ahler geometry with torsion along complex planes in $\bC^{2k}$.
We find that the resulting geometry is K\"ahler with torsion.
The moduli space of these geometries is also investigated.
In addition the applications of these new geometries to M-theory
and sigma models are presented.
In particular, we
 find new solutions of IIA supergravity
 with  the interpretation of
intersecting NS-5-branes at $Sp(2)$-angles
 on a string and show that they preserve $3/32$, $1/8$, $5/32$
and $3/16$
of  supersymmetry. We also show that two-dimensional sigma models
with target spaces the above manifolds have (p,q) extended
supersymmetry. }

\endpage
\pagenumber=2
%\sequentialequations

\chapter{Introduction}

Much insight into the classical and quantum structure of field
theories, strings and M-theory can be obtained by investigating
the classical solutions of these systems.
In particular, the study of
 superstring dualities and M-theory
involves the study of configurations 
that preserve some spacetime
supersymmetry. There is a large 
number of such configurations which
 can be put into two classes. 
The configurations of the first class are 
`elementary'. The other 
class contains configurations that
are constructed by composing or superposing elementary ones.
The elementary configurations typically
preserve $1/2$ of spacetime supersymmetry whereas the composed
ones can preserve $1/2$ or less.

In M-theory, the class of \lq elementary'  configurations
is rather small but they lead to a large number of composite
configurations. These can be constructed using a few simple rules.
In particular, M-theory has two \lq elementary'
 brane configurations: the membrane
and the M-5-brane. Using the M-brane intersection rules,
one is able to construct a large number of M-theory
configurations that have the interpretation 
of intersecting M-branes [\gppkt].
These have found many applications in  
black holes (for a review see [\revone]
and references within)
and Yang-Mills theories [\witten].
Another  application of the above set of 
ideas is to the construction
of a large class of intersecting 
brane solutions of D=11 supergravity theory
starting from the membrane and M-5-brane [\gppkt-\eric].
The first  configurations 
that were constructed preserving some  spacetime supersymmetry
 had the interpretation of {\it orthogonal} 
brane intersections. Later
 it was realized that 
non-orthogonal brane intersections can also preserve some
  spacetime supersymmetry [\leigh-\papasa].
Representing  branes as planes embedded in a 
vector space, supersymmetry
is preserved only if the embeddings are 
chosen in a particular way. In fact
it was found in  [\papas,\west, \spence, \figue] that
 there is a correspondence
 between supersymmetric
intersecting brane configurations and 
calibrations [\calia-\calic]. In this context
the branes that are involved in an 
intersection must lie along calibrated
planes.

In general, it appears that one
can begin from supersymmetric or BPS-like  elementary
configurations and construct
many others 
by superposing them along 
particular planes in a higher-dimensional
vector space. The properties of the resulting configuration depend
on the planes that are chosen. A typical choice of planes are those
associated with calibrations. These lead to composite configurations that
preserve some  spacetime supersymmetry or satisfy a BPS-like
condition. These methods apply to various theories. For example,
the relation between calibrations and Yang-Mills instanton-like
configurations in various dimensions has been examined in [\donaldson].
Later in [\gpt], a class of the $4k$-dimensional instanton solutions of
[\kent] were interpreted as four-dimensional 
instantons superposed along quaternionic
lines in $\bH^{k}$. 
Another application  is in the context of
hyper-K\"ahler geometry. The  $4k$-dimensional toric hyper-K\"ahler
geometries constructed in [\ptg] can be thought of as
superpositions of four-dimensional Taub/NUT geometries along three-planes
in $\bR^{3k}$. Similarly, the $4k$-dimensional hyper-K\"ahler 
geometries with torsion (HKT)
of [\kelly] can be constructed as superpositions of the 
four-dimensional HKT
geometry associated with the NS-5-branes along planes in $\bH^k$.
Both the toric hyper-K\"ahler and HKT geometries in $4k$ dimensions
have found applications in the study of the moduli spaces of 
BPS monopoles [\yi,\gwgnm]  
and five-dimensional black-holes [\kelly], respectively.

In this paper, we shall investigate  various superpositions
of the geometry associated with the NS-5-brane. The non-trivial
part of the metric of the NS-5-brane in the string frame defines
a four-dimensional HKT geometry.
We shall first superpose this four-dimensional HKT geometry
using quaternionic  maps $\tau$ from $\bH^k$ into $\bH$. This will 
lead to new HKT geometries
in $4k$-dimensions which generalize those of [\kelly]. We shall then 
describe several special cases
by choosing different coefficients  for the maps  $\tau$. We shall show that
the eight-dimensional HKT geometries can be constructed
using quaternionic calibrations.
In addition, we shall describe the moduli space of the HKT geometries.
We shall show that it is constructed from the canonical bundle
over the Grassmannian of quaternionic lines in $\bH^k$.
The eight-dimensional HKT geomeries $(k=2)$ have
already been used to construct solutions of IIA supergravity
with the interpretation of NS-5-branes intersecting on a string [\ptg, \ptb].
Here we shall show that these solutions preserve the 
fractions $3/32$, $1/8$, $5/32$ and
$3/16$ of spacetime supersymmetry depending on the 
choice of coefficients
for the  maps.
Then we shall investigate other superpositions of the four-dimensional
HKT geometry associated with the NS-5-brane using  holomorphic  maps.
This will lead to the construction of new K\"ahler geometries with torsion
(KT) in every even dimension. We shall also find the moduli space of these
KT geometries and discuss some of their applications in the context
of supersymmetric two-dimensional sigma models. In particular we find
new two-dimensional (2,0)-supersymmetric
 sigma models. We shall also
 construct new
two-dimensional sigma models with 
(4,2) and (2,2) supersymmetry.

This paper has been organized as follows: 
In section two, we  give the construction of $4k$-dimensional
HKT geometries.
In section three, we  investigate the structure of their moduli space 
 and explain the
relation to calibrations. In section four, we  present 
the applications of these
geometries to intersecting NS-5-branes. In section five, we give
the construction of new KT geometries in every even dimension and describe
their moduli space. In section
six, we  present the application to two-dimensional supersymmetric
sigma models. In section seven,
we  describe some other ways to superpose the four-dimensional
HKT geometry and in section eight we  give
our conclusions. Finally in appendix A, we construct 
five-brane solutions that preserve less supersymmetry.

\chapter{HKT geometries in 4k dimensions}

\section{HKT geometries}

A $4k$-dimensional hyper-complex manifold $M$ with a 
tri-hermitian metric $g$ admits
an HKT structure if the holonomy of one of 
the metric connections
$$
\nabla^{(\pm)}=\nabla\pm H
\eqn\conpm
$$
is $Sp(k)$ with respect to the hyper-complex 
structure, where $\nabla$ is the 
Levi-Civita connection and 
$H(X,Y,Z)=g(X, H(Y,Z))$ is a 
{\it closed} three-form on $M$. 

For later use, we shall give the 
definition of a K\"ahler manifold with torsion (KT).
Let $(M, g,H)$ be  a hermitian manifold
  with metric $g$, complex structure $J$ and a
closed 3-form $H$. Then $M$ admits a 
K\"ahler structure with torsion if the complex structure $J$ is
 covariantly constant with respect to
one of the connections
$\nabla^{(\pm)}$ defined as in \conpm.

Let $\omega_1,
\omega_2$ be the K\"ahler forms of the 
complex structures
$J_1, J_2$, {i.e.}
$$
\omega_1(X,Y)=g(X,J_1Y)
\eqn\mmone
$$
and similarly for $\omega_2$.
The covariant 
constancy condition of the complex
structures  with respect to one of the 
connections $\nabla^{(\pm)}$, say the
$\nabla^{(+)}$ one,  can be replaced [\howec] by
$$
\eqalign{
d\omega_1-2i_1H&=0
\cr
d\omega_2-2i_2H&=0\ ,}
\eqn\hktc
$$
 where 
$i_1, i_2$ are the inner derivations
with respect to the complex structures 
$J_1$ and $J_2$, respectively\foot{Our conventions for a
 p-form are $\omega_p={1\over p!}
\omega_{M_1,\dots, M_p} dx^{M_1}\wedge \dots \wedge dx^{M_p}$ 
and the action of the inner
derivation of the complex structure $J_1$ on $dx^M$ is 
$i_1(dx^M)= J_1{}^M{}_N dx^N$.}.
We remark that the two conditions in
\hktc\ imply a similar condition for the
 third complex structure $J_3$. To find
the associated conditions for the 
$\nabla^{(-)}$ connection, we simply set
$H\rightarrow -H$ in \hktc.

An example of a four-dimensional HKT geometry is 
$$
\eqalign{
d\os^2&={ |dq|^2 \over |q|^2} 
\cr
\of_3&={1\over 3!}{\rm Re}\big( {d\bar q\wedge dq\wedge (d\bar q q 
-\bar q dq)\over 2 |q|^4}\big)\ ,}
\eqn\hktonea
$$
on $\bH-\{0\}=\bR\times S^3$,
where $q\in \bH$ is a quaternion\foot{This geometry is closely 
related to the NS-5-brane solution of 
IIA supergravity in the string frame.}. The metric is
complete and $\of_3$ is the volume form on $S^3$.
In fact this geometry admits two HKT structures. One
is with respect to the pair $(\nabla^{(+)}, J_r)$,
where the complex structures are defined by right-multiplication
with the imaginary unit quaternions $i,j,k$ as 
$$
\eqalign{
J_1&:\quad dq\rightarrow -dq\,i
\cr
J_2&:\quad dq\rightarrow -dq\,j
\cr
J_3&:\quad dq\rightarrow -dq\,k\ .}
\eqn\rcom
$$
The other HKT structure is with respect to the pair 
$(\nabla^{(-)}, I_r)$, where
$$
\eqalign{
I_1&:\quad dq\rightarrow i\,dq
\cr
I_2&:\quad dq\rightarrow j\,dq
\cr
I_3&:\quad dq\rightarrow k\,dq\ .}
\eqn\hheight
$$
In what follows, we shall appropriately 
superpose this four-dimensional
 HKT geometry to construct new  HKT  and 
KT geometries. We shall then present
some of the applications of these geometries 
in M-theory and supersymmetric sigma models.

\section{HKT geometries and quaternionic maps}

We consider the   maps
$$
\tau:\quad \bH^k\rightarrow \bH
\eqn\ggsix
$$
 such that
$$
q\equiv \tau(u)=p_i u^i -a
\eqn\qlinear
$$
where $(u^1,\dots,u^k)\in \bH^k$, $q\in \bH$, 
and $\{p_1,\dots,p_k; a\}$ are quaternions
that parameterize the  map. This map
has $4k$ {\sl rotational}  $\{p_1,\dots,p_k\}$
and four {\sl translational} $\{a\}$  parameters.

The $4k$-dimensional HKT geometry $(ds_{(4k)}^2,H)$ which arises from 
superposing the four-dimensional HKT geometry above, \hktonea, is 
$$
\eqalign{
ds_{(4k)}^2&=ds^2_\infty+ \sum_\tau \mu(\tau) \tau^* d\os^2 
\cr
H&=\sum_\tau \mu(\tau) \tau^*\of_3 \ ,}
\eqn\hkttwoo
$$
where $ds^2_\infty$ is a flat  metric on $\bH^k$, 
$\mu(\tau)$ are real constants and the
sum is over different choices of the  map 
$\tau$. Writing the $ds_{(4k)}^2$ and $H$
explicitly, we have 
$$
\eqalign{
 ds^2_{(4k)}&=ds^2_\infty+\sum_{\{p,a\}} \mu(\{p,a\}) {|
p_i du^i|^2\over| p_i u^i -a|^2} \ .
\cr
H&={1\over 3!}\sum_{\{p,a\}} \mu(\{p,a\}){\rm Re}\big[ 
{d\bar u^i \bar p_i\wedge  p_jd u^j\wedge \big(d\bar
u^k \bar p_k ( p_\ell u^\ell -a) -
(\bar u^m \bar p_m-\bar a)  p_n du^n \big)\over 2| p_i
u^i -a|^4}\big]}
\eqn\hkttwooo
$$
The geometry  \hkttwoo\ on $\bH^k-\cup_\tau \tau^{-1}(0)$ 
admits  an HKT structure
 with respect to the connection $\Gamma^{(+)}$ and the 
complex structures
$$
\eqalign{
{\bf J}_1&:\quad du^i\rightarrow -du^i\, i
\cr
{\bf J}_2&:\quad du^i\rightarrow -du^i\, j
\cr
{\bf J}_3&:\quad du^i\rightarrow -du^i\, k\ .}
\eqn\rcomb
$$
To show
this, we first  observe that $H$ in \hkttwoo\ 
is a closed three-form as required. Also, 
 ${\bf J}_1, {\bf J}_2, {\bf J}_3$ are 
integrable complex structures since they are
constant.
 A straightforward computation reveals
 that the metric \hkttwoo\ is hermitian with
respect to ${\bf J}_1, {\bf J}_2, {\bf J}_3$ 
provided that $ds^2_\infty$ is chosen to be
hermitian with respect to these complex structures.  We
remark that it is always possible to find such a flat metric on $\bH^k$.  
It remains to show that \hkttwoo\ satisfies 
 condition \hktc\ with respect to the 
complex structures \rcomb. For this, we  observe
that 
$$
\eqalign{
d\tau {\bf J}_1 &=J_1  d\tau  
\cr
 d\tau {\bf J}_2 &=J_2 d\tau  
\cr
 d\tau {\bf J}_3&=J_3 d\tau \ .}
\eqn\wwonw
$$
Next we remark that
 \hktc\ is linear in the metric and torsion.
 Using the commutativity of the
complex structures with $d\tau$, \hktc\ can be written as 
$$
\sum_\tau \tau^*\big (d\oo_1-2i_1 \of_3\big)=0
\eqn\wwtwo
$$ 
for the first complex structure, where $\oo_1$ is the K\"ahler
form of the complex structure $J_1$ with respect to the
metric $d\os^2$ in \hktonea,  and 
similarly for the other two complex structures.
However, each part of the sum over 
$\tau$ vanishes since the geometry that we are
pulling-back from $\bH-\{0\}$ is an 
HKT geometry. Thus \hktc\ is also satisfied, and so 
\hkttwoo\  admits an HKT structure.

\section{Special Cases}

We shall consider the following three special cases. 

${(i)}:\,$ We introduce
another triplet of  complex structures on $\bH^k$ as follows:
$$
\eqalign{
{\bf I}_1&:\quad du^i\rightarrow i\, du^i 
\cr
{\bf I}_2&:\quad du^i\rightarrow j\, du^i
\cr
{\bf I}_3&:\quad du^i\rightarrow k\, du^i \ .}
\eqn\rcombb
$$
For generic choices of parameters 
$\{p_1,\dots,p_k\}$ the  maps $\tau$ do not
commute with the complex structures 
${\bf I}_1, {\bf I}_2$ and ${\bf I}_3$. However
if the parameters  $\{p_1,\dots,p_k\}$ of
$\tau$ are {\it real numbers}, i.e.
$$
(p_1,\dots,p_k)\in \bR^k\ ,
\eqn\realnumbersa
$$
instead
 of quaternions, then  $d\tau$ also
commutes with the pairs of complex structures 
$(I_1, {\bf I}_1)$, $(I_2, {\bf I}_2)$ and $(I_3, {\bf I}_3)$. As a result
repeating the argument mentioned above,
\hkttwoo\  admits another HKT structure 
with respect to the pair $(\nabla^{(-)}, {\bf I}_r)$
 provided
that the asymptotic metric $ds^2_\infty$ 
is chosen to be hermitian with respect to all complex
structures\foot{There is always such a 
metric on $\bH^k$.}. 
 These 
geometries with two HKT structures
have already been investigated in [\ptg,\kelly].  

${(ii)}:\,$ Another special case is to choose 
the parameters $\{p_1, \dots,p_k\}$ of the  maps $\tau$
 such that 
$d\tau$ commute with only one of the pairs of 
 complex structures $\{(I_r, {\bf I}_r)\}$, say $(I_1, {\bf I}_1)$.
This implies that the parameters are {\it complex} numbers, i.e.
$$
(p_1,\dots,p_k)\in \bC^k\ ,
\eqn\complexnumbersa
$$
and so
$$
p_i=a_i+ib_i
\eqn\wwthree
$$
where $\{a_1, \dots, a_k; b_1, \dots, b_k\}$ 
are real numbers.
For such a choice of  maps, \hkttwoo\  admits an 
HKT structure with respect to  
$(\nabla^{(+)}, {\bf J}_r)$. In addition, 
\hkttwoo\ admits a
 K\"ahler structure with torsion with respect
to $(\nabla^{(-)}, {\bf I}_1)$, provided
the asymptotic metric $ds^2_\infty$ 
is chosen to be hermitian with respect to ${\bf J}_1,{\bf J}_2,{\bf
J}_3$ and ${\bf I}_1$.  
Therefore the holonomy
of $\nabla^{(-)}$ is a subgroup of $U(2k)$.
In fact, we shall show later that the holonomy of
$\nabla^{(-)}$ is a subgroup of $SU(2k)$. The proof of this
statement will be described for $k=2$ when we give the applications
of these geometries to strings but it  generalizes trivially to
any even dimension.

${(iii)}:\,$ For the third special case, we take $k=2$. The rotational
parameters of the  maps $\tau$ are two quaternions $\{p_1, p_2\}$.
The restriction on the rotational parameters is
$$
{\rm Re} \big( \bar p_1 p_2 \big)=0\ ,
\eqn\imquater
$$
i.e. $\bar p_1 p_2\in {\rm Im}\bH$. 
 The resulting eight-dimensional geometry admits an HKT structure
with respect to $(\nabla^{(+)}, {\bf J}_r)$ as in all the above cases.
In addition, it turns out that the holonomy of $\nabla^{(-)}$ is
 ${\rm Spin}(7)$ provided $ds^2_\infty$ is chosen appropriately.
For example one can choose $ds^2_\infty$ to be the Euclidean 
metric, but more general asymptotic metrics are possible.

The reason for the restriction \imquater\ on $\tau$ will 
become clear later. In particular,
using an equivalence relation on the space of parameters of $\tau$ that 
we shall give in the next section, it is easy to see that
the previous two conditions  \realnumbersa\ and \wwthree\ can be rewritten
as 
$$
\bar p_1 p_2\in \bR
\eqn\mone
$$ 
and 
$$
\bar p_1 p_2 \in \bC\ ,
\eqn\mtwo
$$ 
respectively. 
We remark that other similar conditions to that of \imquater\ 
can be imposed on
the rotational parameters of $\tau$. For example, we can set
$$
{\rm Re} \big(i \bar p_1 p_2 \big)=0\ ,
\eqn\imquateri
$$
and similarly for $j$ and $k$. However, the properties of
 HKT geometries that result from these
conditions are similar to those associated with \imquater\ and 
we shall not investigate them
further here.

\chapter{Grassmannians and Calibrations}
\section{The Moduli Space}

The HKT geometry \hkttwooo\ is determined
{\sl solely} by the arrangement of  
quaternionic (k-1)-planes in $\bH^k$ given by
the kernels of the  maps $\tau$.
To see this, we note that the HKT geometries of section (2.2)
are invariant on replacing
each map $\tau$ with parameters $\{p_1, p_2, \dots, p_k; a\}$
with another  map $\hat \tau$ with
 parameters $\{ s p_1 , s p_2 , \dots, s p_k; s a\}$ where $s\in \bH-{0}$.
This is because both the metric and torsion of the
HKT geometry are invariant under a rescaling of the parameters of the
 maps with a quaternion from the left.
Therefore, we define an equivalence relation $\sim $ on the space of
maps $\tau $, so that 
$\hat\tau\sim \tau$ i.e
$$
(\hat p_1, \hat p_2, \dots, \hat p_k; \hat a)\sim (p_1, p_2, \dots, p_k; a)
\eqn\equivrel
$$
if there exists an $s\in \bH-{0}$ for which $ \hat\tau= s \tau$, i.e
$$
(\hat p_1, \hat p_2, \dots, \hat p_k; \hat a)=(s p_1, s p_2, \dots, s p_k; s a)\ .
\eqn\equivaa
$$
The space of equivalence classes of  maps  is 
the bundle space $E(\gamma^{k-1}_1)$ of the canonical quaternionic
line bundle $\gamma^{k-1}_1$    over the Grassmannian
$$
Gr(1, \bH^k)=Sp(k)/Sp(1)\times Sp(k-1)\ .
\eqn\wwfour
$$
The fiber directions of $\gamma^{k-1}_1$  are associated 
with the translation parameters
  whereas the rest are 
associated with the rotational
parameters of the  maps. In particular, $a$ is the fibre
coordinate of $\gamma^{k-1}_1$ and $(p_1, \dots, p_k)$ are the
homogeneous coordinates of $Gr(1, \bH^k)$. Observe that the 
maps that lie within the same equivalence class have the same kernel.
         
The moduli space, ${\cal M}_N(\bH)$, of HKT geometries
 associated with $N$ distinct 
maps $\tau$, i.e.  maps that are not
 equivalent in the  sense of \equivrel,  is
$$ 
{\cal M}_N(\bH)={D_N\times \big(\times^N  E(\gamma^{k-1}_1)-\Delta\big)\over S_N}\ ,
\eqn\modulii
$$
where $D_N$ is a domain in $\bR^N$ which 
parameterizes the scale factor of each  term in
the sum for the metric (and torsion),
$\Delta$ is a diagonal term in $\times^N  E(\gamma^{k-1}_1)$ and $S_N$ 
is the 
permutation group of $N$ points. $D_N$ includes the $(\bR^+)^N$ 
subspace of $\bR^N$.
The moduli
space, 
$\tilde {\cal M}_N(\bH)$, of rotation 
and translation parameters of such HKT geometry is
$$ 
\tilde{\cal M}_N(\bH)={\times^N  E(\gamma^{k-1}_1)-\Delta\over S_N}\ .
\eqn\mthree
$$

We remark that these moduli spaces are different
from the ones that arise in the study of extreme black holes.
The moduli space of N-indistinguishable black holes
is isomorphic to the configuration space of N-indistinguishable 
particles.
The coordinates of the moduli space of extreme black holes
specify their location
and in our terminology consist of translational parameters.
However, as we have seen, the moduli spaces of our configurations
contain also 
{\sl rotational} parameters.

The simplest case to consider is  $N=2$ and $k=2$. The 
diagonal term in this case is
$$
\Delta=\{( x_1,  x_2)\in \times^2  E(\gamma^1_1):
x_1=x_2\}
\eqn\mfour
$$
It is clear that if $x_1=x_2$, then  the metric degenerates, 
ignoring the asymptotic
part, 
to the four-dimensional
HKT one.

It remains to investigate the moduli of these geometries in the
three special cases of section (2.3). 

${(i)}\,$ If the rotational parameters of the  maps
$\tau$ are real numbers, then the equivalence relation is as in
\equivrel\ and \equivaa\ but $s$ in this case is a real number.
Therefore, the space of parameters is the
 bundle space $E(\oplus^4\zeta^{k-1}_1)$, where
$\zeta^{k-1}_1$  is the canonical real line bundle over the
 real projective space
$$
\bR P^{k-1}= {SO(k)\over 
\bZ_2\times SO(k-1)}={S^{k-1}\over \bZ_2}\ .
\eqn\mfive
$$

The moduli space, ${\cal M}_N(\bR)$, of HKT geometries
 associated with $N$ distinct 
maps $\tau$ with real rotational parameters  is
$$ 
{\cal M}_N(\bR)={D_N\times \big(\times^N 
 E(\oplus^4 \zeta^{k-1}_1)-\Delta\big)\over
S_N}\ ,
\eqn\msix
$$
where $D_N$, $\Delta$ and $S_N$ are defined as in \modulii.

${(ii)}:\,$ Now, if the rotational parameters
 of the  maps are complex numbers, the space of parameters
of $\tau$ is the bundle space $E(\xi^{k-1}_1\oplus \xi^{k-1}_1)$, where 
$\xi^{k-1}_1$ is the canonical complex line bundle over the complex 
projective space
$$
\bC P^{k-1}={U(k)\over U(1)\times U(k-1)}\ .
\eqn\mseven
$$
The moduli space, ${\cal M}_N(\bC)$, of HKT geometries
 associated with $N$ distinct 
maps $\tau$ is
$$ 
{\cal M}_N(\bC)={D_N\times \big(\times^N 
 E(\xi^{k-1}_1\oplus \xi^{k-1}_1)-\Delta\big)\over
S_N}\ .
\eqn\meight
$$

${(iii)}:\,$ To investigate the moduli space for the 
third case in section (2.3), we
observe that the relation ${\rm Re} \bar p_1 p_2=0$ is invariant
under the equivalence relation  \equivrel. Then it is easy  to show
that the space of
equivalence classes of linear maps is isomorphic to the
bundle space $E$ of a quaternionic line bundle $\eta$ over
$S^3$.
The moduli space, ${\cal M}_N({\rm Im}\bH)$, of HKT geometries
 associated with $N$ distinct 
maps $\tau$ is then
$$ 
{\cal M}_N({\rm Im}\bH)={D_N\times \big(\times^N 
 E(\eta)-\Delta\big)\over
S_N}\ .
\eqn\mnine
$$

\section{Calibrations and HKT geometries}

In this section we shall see that the new $k=2$  HKT geometries
that we have constructed above are related 
to calibrations in eight dimensions.
 We shall find that  they are ``calibrated''
geometries in the sense that they 
are superpositions of a model four-dimensional geometry along
planes in the contact set of some calibration.
Below we give a short summary of some of the main properties
of calibrations that we shall use here. For a detailed
account of calibrations and their main application to the
construction of minimal surfaces we refer the reader to
[\calia, \calib].

A degree $\ell$ calibration in $\bR^n$ is associated 
with a certain closed $\ell$-form $\phi$
defined on $\bR^n$. In most applications $\phi$
is a constant form on $\bR^n$. In addition, $\phi$ is chosen such that
if the co-volume form $\zeta$ of any $\ell$-plane is evaluated
on $\phi$,  then $\phi(\zeta)\leq 1$. 
We say that an $\ell$-plane with 
co-volume form
$\zeta$ is calibrated by $\phi$, if  $\phi(\zeta)=1$. The set of 
calibrated planes of $\phi$ is a subset of the
Grassmannian  of oriented $\ell$-planes in $\bR^n$, $Gr(\ell, \bR^n)$. 
We shall refer to the set of calibrated planes of $\phi$ as 
 the contact set of the calibration and denote it with
$G_\phi$. In most cases, the contact set is a homogeneous space $G/H$.

A large class of calibrations in $\bR^8$ were investigated in [\calic].
Here we shall use those associated with  constant self-dual forms in $\bR^8$.
We begin by choosing on $\bR^8$ the 
hyper-complex $\{{\bf J}_1, {\bf J}_2, {\bf J}_3\}$  
of section two. Let $\{\omega_{{\bf J_1}}, \omega_{{\bf J_2}}, \omega_{{\bf J_3}}\}$
be the associated K\" ahler forms with respect to the flat metric.
A calibrating 4-form is 
$$
\Phi_J=-{1\over 6}(\omega_{{\bf J_1}}^2 + \omega_{{\bf J_2}}^2 +
\omega_{{\bf J_3}}^2)\ ,
\eqn\mten
$$
with contact set the grassmannian
$$
Gr(1, \bH^2)=Sp(2)/Sp(1)\times Sp(1)=S^4\ ,
\eqn\wwfour
$$
where  $\omega_{{\bf J}_1}^2= \omega_{{\bf J}_1}
\wedge \omega_{{\bf J_1}}$ and similarly for ${\bf J}_2$ and ${\bf J}_3$. Note
that $\Phi_J$ is the $Sp(2)\cdot Sp(1)$-invariant quaternionic four-form associated
with $\{{\bf J}_1, {\bf J}_2, {\bf J}_3\}$.

There are special cases of the above calibration for which the contact set
is a subspace of \wwfour :

$(i):\,$ Let $\Phi_I$ be the quaternionic four-form associated with
$\{ {\bf I}_1, {\bf I}_2, {\bf I}_3\}$ of section two.  A calibration form is
then
$$
\Theta = {1\over 2}(\Phi_I + \Phi_J)\ ,
\eqn\aaone
$$
with contact set given by
$$
G_\Theta=S^1\ .
\eqn\aatwo
$$

$(ii):\,$ Let $\omega_{{\bf I}_1}$ be the K\"ahler form of
the ${\bf I}_1$ complex structure of section two with 
respect to the flat metric.
A calibration
form is then
$$
\Lambda={1\over5} \omega^2_{{\bf I}_1}+{3\over5} \Phi_J\ ,
\eqn\aathree
$$
and the contact set is
$$
G_{\Lambda}=SU(2)/S\big(U(1)\times U(1)\big)=S^2\ .
\eqn\aafour
$$

$(iii):\,$ Let $\Omega$ be a certain self-dual $Spin(7)$-invariant 
four-form. A calibration form is
$$
\Psi={1\over4} \Omega +{3\over 4} \Phi_J\ ,
\eqn\aafive
$$
with contact set given by
$$
G_\Psi={Sp(1)\times Sp(1)\over Sp(1)}=S^3\ .
\eqn\aasix
$$ 

We proceed now to establish the correspondence of 
our HKT geometries with the above calibrations.
As shown in the previous section our HKT geometries are determined
by the maps $\tau$. The kernels of the maps $\tau$ are 
$$
p_i u^i-a=0\ ,
\eqn\qline
$$
where $i=1,2$. 
We find that if the rotational
parameters $\{p_1, p_2\}$ are quaternions, 
then the kernels of maps $\tau$
are calibrated by $\Phi_J$. To see this, 
we observe that their tangent
spaces are stabilized by the action of  
${\bf J}_1,{\bf J}_2$ and hence also by
${\bf J}_3$, and so they are quaternionic 
lines in $\bH^2$. Thus 
they are calibrated by $\Phi_J$.

Let us investigate now the special cases 
where we restrict the rotational
parameters to be (i) real, (ii) complex 
and (iii) satisfy ${\rm Re} (\bar p_1 p_2)=0$.

$(i):\,$ If the rotational parameters are 
real numbers, then the tangent spaces
of the kernels of $\tau$ are stabilized by 
the additional hyper-complex structure
$\{ {\bf I}_1, {\bf I}_2, {\bf I}_3\}$. So 
they are calibrated simultaneously by
$\Phi_J$ and $\Phi_I$, and thus also by $\Theta$.

$(ii):\,$  Choosing, as in section (2.3), 
$p_i=a_i+i b_i$, we observe that  the tangent
spaces of the kernels of $\tau$ are stabilized by 
${\bf I}_1$. So they are calibrated simultaneously by
$\Phi_J$ and ${1\over 2}\omega^2_{{\bf I}_1}$, 
and thus also by $\Lambda$. 

$(iii):\,$  Restricting, as in section (2.3), 
${\rm Re} (\bar p_1 p_2)=0$, we find that  
 the kernels of $\tau$ are  calibrated by the 
self-dual $Spin(7)$-invariant form $\Omega$
with non-vanishing components
$$
\Omega_{1234}=\Omega_{1265}=\Omega_{1287}=
\Omega_{1375}=\Omega_{1476}=\Omega_{1368}=\Omega_{1485}=1\ .
\eqn\aaseven
$$
To see this, we evaluate the co-volume form $\eta$ of 
the kernels on $\Psi$ and find
$$
\Psi(\eta)={ (\bar p_1 p_1+ \bar p_2 p_2)^2- 2 
({\rm Re}( \bar p_1 p_2))^2\over (\bar p_1 p_1+
\bar p_2 p_2)^2}\ .
\eqn\aaeight 
$$
So $\Psi(\eta)=1$, iff ${\rm Re} (\bar p_1 p_2)=0$.
It then follows that the kernels of the maps $\tau $ are also
calibrated by $\Omega $ iff ${\rm Re} (\bar p_1 p_2)=0$.

In all the above cases, the only planes 
that are calibrated by the corresponding
four-forms are those given by the 
associated maps $\tau$ (see [\calic]).
This explains why the moduli spaces of 
the eight-dimensional HKT geometries can be
constructed using projective spaces which
 are isomorphic to the
contact sets of the above calibrations. It would be of 
interest to extend this result
to $4k$, $k>2$, dimensions.

\section{Angles}

The asymptotic angles between the 
planes defined by the kernels of two 
maps $\rho$ and $\tau$ can be 
computed using the normal vectors
${\bf n}$ and ${\bf m}$ of these 
planes,
respectively\foot{Each plane has 
four such normal vectors.}. 
These are given by
$$
\eqalign{
<{\bf n}, X>&=d\rho(X)
\cr
<{\bf m}, X>&=d\tau(X)}
\eqn\aanine
$$
where the inner product $<\cdot,\cdot>$ is with respect to the 
metric at infinity. The angles-matrix is then defined as
$$
A={<{\bf n},{\bf m}>\over |{\bf n}| |{\bf m}|}
\eqn\aaten
$$
where $|{\bf n}|^2=<{\bf n},{\bf n}>$ and similarly for $|{\bf m}|$.  

Choosing as asymptotic metric the 
flat metric on $\bH^k-\cup_\tau \tau^{-1} (0)$,
we find that
$$
A={ p_i \bar q_j \delta^{ij}\over
 {\sqrt {\delta^{ij}  p_i \bar p_j}} {\sqrt
{\delta^{k\ell}
 q_k \bar q_\ell}}}
\eqn\bbone
$$
where $\{p_1, \dots, p_k\}$ 
and $\{q_1, \dots, q_k\}$ are the rotational
parameters of the maps $\rho$ and $\tau$, respectively. The explicit
expression of the angles-matrix for a general asymptotic metric
is a straightforward generalization of that given in [\ptb] for 
the intersecting five-branes and so it will not be presented here.

The angles-matrix $A$ depends on the choice of representative
 in the equivalence relation \equivrel. 
In particular, let $\rho\rightarrow e\rho$
and $\tau\rightarrow \ell\tau$. Then
$$
A\rightarrow { e \over |e|}A\bar {\ell \over |\ell|}\ ,
\eqn\bbtwo
$$
where $e,\ell\in \bH$. Therefore it depends on some of the data of
the parameterization of the  maps $\tau$ unlike the underlying
HKT geometry which depends only on the planes defined by the kernels
of these  maps.

\chapter{Applications}

\section{Intersecting IIA five-branes on a string }

The eight-dimensional HKT geometries found in the previous section
admit a brane interpretation and lead to the supergravity solutions 
constructed in [\ptb].  To relate the two, we  
express the quaternionic coordinates $\{u^i; i=1,2\}$ used
 in the previous section in
terms  of the real coordinates $\{x^{i\mu}; i=1,2; \mu=0,1,2,3\}$ 
used in [\ptb] as $u^i=
x^{i0}+ix^{i1}+j x^{i2}+k x^{i3}$. Let 
$(ds^2_{(8)}, H_{(8)})$ be the metric
and closed three-form of the HKT geometry. 
The IIA supergravity solution
in the string frame  is
$$
\eqalign{
ds^2&=ds^2(\bE^{(1,1)})+ ds^2_{(8)}
\cr
H&=H_{(8)}
\cr
e^{8\phi}&=g_{(8)}\ ,}
\eqn\fbsol
$$
where $ds^2$ is the ten-dimensional supergravity metric,
 $\phi$ is the dilaton, $H$ is
the NS$\otimes$NS three-form field strength and $g_{(8)}$ is the
 determinant of the HKT metric $ds^2_{(8)}$.
This solution has the interpretation of  NS-5-branes intersecting
on a string. The positions of the NS-5-branes are determined by the
kernels of the  maps $\tau$.

%\bigskip
%\epsfscale=1000
%\centerline{\epsfbox{branes.eps}}
%\vskip 0.2cm
%\noindent{\bf Fig 1: Intersecting NS-5-branes on a String}\ \ \
%The lines  denote the location of some of
% the four-planes in $\bH^8$ associated
%with the kernels of the maps $\tau$ of our
%solution. These lines  also denote the 
%locations of the 5-branes of the
%configuration.
%\noindent

We proceed to determine the proportion of spacetime supersymmetry
preserved by the above solution. Let $\Gamma$ be the Levi-Civita
connection of the supergravity metric. The IIA Killing spinor
equations in the string frame  are
$$
\eqalign{
\nabla^{(+)}_M\epsilon&=0\ ,
\cr
\nabla^{(-)}_M\eta&=0
\cr
\big(\Gamma^M\partial_M\phi- {1\over6} H_{MNP} \Gamma^{MNP}\big)\epsilon&=0
\cr 
\big(\Gamma^M\partial_M\phi+ {1\over6} H_{MNP} \Gamma^{MNP}\big)\eta&=0\ ,}
\eqn\killspin
$$
where $\epsilon$ and $\eta$ are the chiral
 and  anti-chiral Majorana-Weyl Killing
spinors, respectively, 
$\nabla^{(\pm)}$ are the covariant
 derivatives of the connection
$$
\Gamma^{(\pm)}{}^M{}_{NP}= \Gamma^M_{NP}\pm H^M{}_{NP}\ ,
\eqn\conpm
$$
and $M, N, P=0, \dots, 9$.
It is clear that the first two Killing spinor
 equations have solutions provided
that the decompositions of the Majorana-Weyl
 (chiral and anti-chiral) representations of 
${\rm Spin}(1,9)$ have singlets
under the holonomy groups of the connections 
$\nabla^{(\pm)}$. The number of linearly independent
spinors determines the proportion of 
 supersymmetry preserved by the solution.

The holonomies of the $\nabla^{(\pm)}$
connections of the solution \fbsol\ are those of the associated connections
of the HKT geometry in \fbsol. It follows from the
results in [\ptg] that the holonomy of the connection
$\nabla^{(+)}$ is exactly  
$Sp(2)$ for a generic choice of  maps $\tau$.
Moreover, we have verified by an explicit computation of
the curvature tensor, that for  a generic choice
of  
maps $\tau$, the holonomy group of 
 $\nabla^{(-)}$  is exactly $SO(8)$.
Therefore, the only solution
 of the second Killing spinor
equation is the trivial one, i.e. $\eta=0$.  
With the above choice of dilaton the third 
Killing spinor equation is 
 satisfied without any additional
conditions. In fact, the first Killing
 spinor equation implies the third
provided the holonomy of the connection $\nabla^{(+)}$ is a
subgroup of $SU(4)$, [\strominger, \callan]. In our context
the condition for $SU(4)$ holonomy can be expressed, in a
coordinate system relative to which the associated complex
structure is constant, as a condition on the metric,
$$
g^{ab}\partial_{a} g_{bc}={1\over4} g^{ab}
\partial_{c} g_{ab}\ ,
\eqn\keyrel
$$
where $a,b,c=1,\dots, 8$. This is precisely the 
condition needed to satisfy the third Killing spinor
equation.

To determine the proportion of  
supersymmetry preserved by the
solution, we have to find the decomposition
of the Majorana-Weyl representation of ${\rm Spin}(9,1)$
under  the holonomy group of $\nabla^{(+)}$. 
 We first decompose
the Majorana-Weyl representation of
${\rm Cliff}(9,1)$ under the ${\rm Spin}(8)$ 
subgroup of ${\rm Spin}(9,1)$ as
$$
{\bf 16}\rightarrow {\bf 8}_s\oplus {\bf 8}_c\ ,
\eqn\mntwo
$$
where ${\bf 8}_s$ is the spinor representation
of ${\rm Spin}(8)$ and ${\bf 8}_c$ is its conjugate.
Next we decompose ${\bf 8}_s$ and ${\bf 8}_c$ 
under $Sp(2)\subset {\rm Spin}(8)$ as
$$
\eqalign{
{\bf 8}_s&\rightarrow {\bf 5}\oplus {\bf 1}\oplus {\bf 1}\oplus {\bf 1}
\cr
 {\bf 8}_c&\rightarrow {\bf 4}\oplus {\bf 4}\ .}
\eqn\mnthree
$$
Note that there are three singlets 
in the decomposition of ${\bf 8}_s$. 
Hence, we conclude that our solution has three Killing spinors 
and therefore
it preserves $3/32$ of  supersymmetry.

\section{Special Cases}

There are three special cases of intersecting five-brane  
configurations to consider. 
These are associated with 
eight-dimensional HKT geometries
for which the rotational parameters $\{p_1, p_2\}$ of the 
  maps $\tau$ are (i) real, (ii) complex
and (iii) satisfy ${\rm Re} (\bar p_1 p_2)=0$. In  case (i), the solution 
has already been investigated in [\ptg].
It has been found that the configuration admits six Killing spinors
and therefore preserves $3/16$ of spacetime supersymmetry.  This 
analysis will not be repeated here.

Next let us take the rotational parameters of the maps
to be complex numbers.
 The analysis
of the first and third Killing spinor
 equations in \killspin\ is as before.
This implies that our solutions admit
  three Killing spinors
associated with the Majorana-Weyl
 (sixteen component) supersymmetry parameter
$\epsilon$.  The analysis of
the second and fourth Killing spinor
 equations associated with 
the other Majorana-Weyl supersymmetry
 parameter $\eta$ of opposite chirality
is different. For this, we note that
these HKT geometries also admit a KT structure
with respect to the pair $(\nabla^{(-)}, {\bf I}_1)$.
This implies that the holonomy of the connection $\nabla^{(-)}$
is a subgroup of $U(4)$. In fact, it turns out that it is a subgroup
of $SU(4)$. One way to show this is to verify that the curvature
$R^{(-)}$ of $\nabla^{(-)}$ takes values in the Lie algebra
of $SU(4)$. However since the complex structure ${\bf I}_1$ is
constant in the natural coordinate
 system that we have introduced,
this condition
 on the curvature is implied by
$$
{\rm Tr}\big(\Gamma^{(-)}_{a} {\bf I}_1\big)=0\ .
\eqn\sunhol
$$
After some computation, one can verify that \sunhol\ holds
using  \keyrel.  The fourth
Killing spinor equation then follows without further
restrictions.

To determine the proportion of  
supersymmetry preserved by the
solution, we have to find the decomposition
of the Majorana-Weyl representation of ${\rm Spin}(9,1)$
under  the holonomy group $SU(4)$ of $\nabla^{(-)}$. 
 We first use the decomposition 
${\bf 16}\rightarrow {\bf 8}_s\oplus {\bf 8}_c$
of the Majorana-Weyl representation of
${\rm Spin}(1,9)$ under
 ${\rm Spin}(8)$ as in \mntwo.
Next we decompose ${\bf 8}_s$ and ${\bf 8}_c$ under $SU(4)$ as
$$
\eqalign{
{\bf 8}_s&\rightarrow {\bf 6}\oplus {\bf 1}\oplus {\bf 1}
\cr
 {\bf 8}_c&\rightarrow {\bf 4}\oplus \bar{\bf 4}\ .}
\eqn\ttone
$$
Hence, we conclude that our solution has 
two additional Killing spinors 
associated with the supersymmetry 
parameter $\eta$. Therefore, the IIA
intersecting five-brane
configurations  given by the HKT 
geometries with complex rotational parameters have 
five Killing spinors and therefore
 preserve $5/32$ of spacetime supersymmetry.

It remains to investigate the intersecting five-brane configurations
associated with an eight-dimensional HKT metric for which the rotational
parameters of the  maps satisfy ${\rm Re} (\bar p_1 p_2)=0$. The holonomy
of the $\nabla^{(+)}$ connection is $Sp(2)$ as in the previous
cases. The holonomy of  $\nabla^{(-)}$ is $Spin(7)$ provided
that $ds^2_\infty$ is chosen appropriately. For this, we have verified 
after some computation that
$$
{1\over 2} \Omega_{\hat a \hat b}{}^{\hat c\hat d} 
R^{(-)}_{\hat c\hat d}= -R^{(-)}_{\hat a\hat b}\ ,
\eqn\bbfive
$$
where $\Omega$ is the $Spin(7)$ invariant form given 
in section (3.2) and $\hat a, \hat
b, \hat c, \hat d=1,\dots, 8$
are tangent space indices. As remarked previously 
we can choose an Euclidean asymptotic metric on the 
eight-dimensional space transverse to the string, although
more general flat tri-hermitian (relative to ${\bf J}_1,{\bf J}_2,{\bf
J}_3$) asymptotic metrics are possible. We shall not discuss
this further here.
Under the decomposition of the Majorana-Weyl
 representation of ${\rm Spin}(9,1)$ under $Spin(7)$,
we find an additional singlet. Furthermore, 
the last Killing spinor equation
in \killspin\ ,i.e the dilatino equation associated with
the $\nabla^{-}$ connection, is satisfied for this spinor. 
This follows from
$$
\partial_{\hat d} \phi =-{1\over 6}\Omega^{
\hat a\hat b \hat c}{}_{\hat d} H_{\hat a\hat b\hat c}  
$$
which can be verified after some careful computation. In the
above $H_{\hat a\hat b\hat c}$ correspond to the tangent
frame components of the NS 3-form field strength. We thus
conclude that this solution preserves $1/8$ of supersymmetry.

We summarize some of our results of the previous two 
 sections  in the table below. 
 \vskip 0.3cm
$$
\vbox{\settabs 5\columns 
\+  $\bar p_1 p_2$& {$\nabla^{(+)}$} & {$\nabla^{(-)}$}&{\rm Susy}&{\rm Contact Set}
\cr
\+{$\bR$}&$Sp(2)$& $Sp(2)$& $~{3\over 16}$& \qquad $S^1$\cr
\+ {$\bC$}&$Sp(2)$& $SU(4)$&$~{5\over 32}$& \qquad $S^2$\cr
\+{${\rm Im} \bH$}& $Sp(2)$& $Spin(7)$&$~{1\over 8}$& \qquad $S^3$\cr  
\+ $\bH$ &$Sp(2)$& $SO(8)$  &$~{3\over 32}$& \qquad $S^4$\cr}
 $$
\vskip 0.2cm
\noindent{\bf Table 1: Intersecting NS-5-branes on a string}\ \
This table contains (i)
the type of rotational parameters $\{ p_1, p_2\}$ of the eight-dimensional 
HKT geometries, (ii) and (iii) the holonomy
of the  associated $\nabla^{(+)}$ and $\nabla^{(-)}$ 
connections, respectively, (iv) the fraction of supersymmetry
preserved by the corresponding intersecting 
NS-5-brane configurations and (v) the contact sets of the associated calibrations.
\noindent

\chapter{KT geometries in diverse dimensions} 

\section{Holomorphic maps and KT structures}

New KT geometries  can be constructed by
superposing the  four-dimensional HKT geometry \hktonea\
using holomorphic  maps.  Let $J$ be one of the complex
structures on $\bE^4$ associated with an anti-self dual two-form. 
Using $J$, we identify $\bE^4$ 
 with $\bC^2$. The
HKT geometry in \hktonea\ is now rewritten as
$$
\eqalign{
d\os^2&={|dz|^2\over |z|^2}
\cr
\of_3&={\delta_{\alpha \bar \beta} \delta_{\gamma \bar\epsilon}\bar
z^{\bar\epsilon} dz^\alpha\wedge
 d\bar z^{\bar\beta}\wedge  dz^\gamma\over
|z|^4}+ {\rm c.c. }}
\eqn\comhkt
$$
where $\{z^1,z^2\}$ are the complex coordinates on $\bC^2$. 

To proceed, we identify $\bE^{4k}=\bC^{2k}$ and introduce
 the 
 holomorphic maps
$$
\tau:\quad \bC^{2k}\rightarrow \bC^2
\eqn\ttthree
$$ 
such that
$$
z^\alpha\equiv \tau^\alpha(w)=p^\alpha{}_{i\beta} w^{i\beta}-a^\alpha
\eqn\ttfour
$$
where $\{w^{i\beta}; i=1,\dots, k; \alpha=1,2\}$ 
are holomorphic coordinates of
$\bC^{2k}$ and $\{p_1,\dots, p_{k};a\}$ 
are the {\sl complex} parameters of the 
map $\tau$. These  maps $\tau$ have 
$8k$ rotational
parameters $\{p_1,\dots, p_{k}\}$  and 
four translational parameters $\{a^\alpha; \alpha=1,2\}$. 
Following the construction of $4k$-dimensional HKT geometries 
in section two, we find a 
new geometry on
$\bC^{2k}-\cup_\tau \tau^{-1}(0)$ by pulling-back
\comhkt\ with generic  maps $\tau$ and then 
summing up over the different choices of $\tau$.
The resulting geometry is
$$
\eqalign{
ds^2&=ds^2_\infty+\sum_{\{p,a\}} \mu(\{p,a\}) {|d(pw)|^2\over |pw-a|^2}
 \cr
H&=\sum_{\{p,a\}} \mu(\{p,a\}){ 
\big[ \delta_{\alpha \bar \rho } \delta_{\beta\bar
\gamma} (\bar p^{\bar \rho}_{i\bar \mu} 
\bar w^{i\bar\mu}- \bar a^{\bar \rho} )
p^\alpha_{j\nu} p^\beta_{k\sigma} 
\bar p^{\bar\gamma}_{\ell \bar\tau} dw^{j\nu}
\wedge dw^{k\sigma} \wedge d\bar w^{\ell \bar \tau}\big] 
\over |pw-a|^4}  +{\rm c.c.}\ ,}
\eqn\ktgeom
$$
where $ds^2_\infty$ is a constant metric on 
$\bC^{2k}$.

It turns out that \ktgeom\
admits a KT structure. To see this, we
 introduce a complex structure ${\bf J}$ on
$\bC^{2k}-\cup_\tau \tau^{-1}(0)$ as 
$$
{\bf J }:\quad dw^{i\alpha}\rightarrow dw^{i\alpha} i\ .
\eqn\ttfive
$$
This complex structure is integrable 
since it is constant. The metric in \ktgeom\
is  hermitian with respect to ${\bf J}$ 
provided that $ds^2_\infty$ is also hermitian
relative to this complex 
structure\foot{It is always possible to choose
$ds^2_\infty$ to be hermitian with 
respect to a complex structure.}.
So it remains to show that \ktgeom\ satisfies \hktc\ with respect
to   ${\bf J}$.  
The proof of this 
is similar to that given for the HKT case in section two. The 
key point is that  the  map $\tau$ is
holomorphic and so $d\tau$ commutes 
with the complex structures $J$ and ${\bf J}$ 
on $\bC^2$ and $\bC^{2k}$,
respectively  .

As we have seen, the method described above has led to the 
construction of KT geometries
in $4k$ dimensions. A simple modification of this method
can also lead to new KT geometries in all even dimensions.
For this, we 
choose maps $\tau:\, \bC^{2k}\rightarrow \bC^2$
which restrict to a subspace $V$ of $\bC^{2k}$, i.e.
$$
\tau: \quad V\subset \bC^{2k}\rightarrow \bC^2\ .
\eqn\bbsix
$$ 
 The vector space $V$ is
complex with complex structure ${\bf J}_V$ 
induced by ${\bf J}$  of $\bC^{2k}$.
Repeating the procedure for constructing KT
 geometries above,
we find a new KT geometry on 
$V-\cup_\tau \tau^{-1}(0)$. In particular,
$J d\tau= d\tau {\bf J}_V$ is satisfied 
because $J d\tau= d\tau {\bf J}$.
Since we are allowed to choose any complex 
subspace $V$ of $\bC^{2k}$
and so subspaces with dimensions which are
 not multiple of four, this
construction will give new KT geometries 
in all even dimensions.

\section{Moduli space and angles}

The  maps $\tau$ used in the previous section
for the construction of KT geometries are parameterized by 
$\{p_1,\dots, p_k; a\}$, where $\{p_1,\dots, p_k\}$ are $2\times 2$
complex matrices and $\{a\}$ is a complex two-vector.
Let $G$ be the group of $2\times 2$ complex matrices such that
$$
g^\dagger g=r {\bf 1}
\eqn\ccone
$$
for some $r\in \bR-{0}$.
Then, it is straightforward to see that the
 maps $\tau$ and $\hat\tau$ with
parameters $\{p_1,\dots, p_k; a\}$ and 
$\{\hat p_1,\dots, \hat p_k; \hat a\}$, respectively, that
satisfy
$$
(\hat p_1,\dots, \hat p_k; \hat a)=(g\,p_1,\dots, g\,p_k; g\,a)\ ,
\eqn\comequiv
$$
where $g\in G$, lead to the same KT geometry. 
So we define the equivalence
relation on the space of  maps
$$
\tau\sim \hat\tau\ ,
\eqn\moreequiv
$$
iff there is $g\in G$ such that their 
parameters are related as in \comequiv\ or
simply $\hat\tau=g\tau$.
The group $G$ is isomorphic to 
$(\bR-\{0\})\times U(2)$.    To
see this, let us first take $r>0$. In this case,
$$
\eqalign{
\Phi:\quad &G\rightarrow \bR^+\times U(2)
\cr 
	&g\rightarrow (|{\det g}|^{{1\over2}}, 
|{\det g}|^{-{1\over2}} g)}
\eqn\cctwo
$$
with inverse
$$
\eqalign{
\Phi^{-1}:\quad \bR^+&\times U(2)\rightarrow G
\cr
(r,& U)\rightarrow rU\ ,}
\eqn\ccthree
$$
is a group isomorphism. Finally if $r<0$, it is 
easy to show that
$G$ is diffeomorphic to $\bR^-\times U(2)$.

Let us denote with $E$ the set  of equivalence classes of 
 maps. It is clear that,
unlike  the quaternionic case in sections of
(2.1)-(2.3), the above KT
geometries are not only specified 
by the kernels of the  maps $\tau$ in $\bC^{2k}$
but in fact also depend on the actual parameterization
of the maps\foot{For this reason, there is not a direct 
relation between these KT geometries and K\"ahler calibrations.}.
The moduli space ${\cal M}_N$ of KT geometries 
associated with N-distinct
 maps $\tau$, i.e.  maps that 
are not equivalent with respect
to the above relation, is
$$
{\cal M}_N={D_N\times\big(\times^N E-\Delta\big)\over S_N}\ ,
\eqn\ccfour
$$ 
where $D_N$ is a domain in $\bR^N$ associated 
with the conformal
factors of every term in the sum over 
the  maps in \ktgeom\ and $\Delta$ is
the diagonal.

The asymptotic angles between the 
planes defined by the kernels of two  holomorphic
maps $\rho$ and $\tau$ can be 
computed using the normal vectors
${\bf n}$ and ${\bf m}$. We define the normal vectors
and the angles matrix as in the quaternionic case. The angle
matrix is 
$$
A={<{\bf n},{\bf m}>\over |{\bf n}| |{\bf m}|}
\eqn\ccfive
$$
where $<\cdot,\cdot>$ is the inner product with respect to
the asymptotic metric and $|{\bf n}|^2=<{\bf n},{\bf
n}>$, and similarly for
$|{\bf m}|$.  

Choosing as asymptotic metric the 
flat metric on $\bC^{2k}-\cup_\tau \tau^{-1} (0)$,
we find that
$$
A={ p_i  q\dagger_j \delta^{ij}\over 
{\sqrt {\delta^{ij}  p\dagger_i p_j}} {\sqrt
{\delta^{k\ell}
 q^\dagger_k q_\ell}}}
\eqn\ccsix
$$
where $\{p_1, \dots, p_k\}$ 
and $\{q_1, \dots, q_k\}$ are the rotational
parameters of the maps $\rho$ and $\tau$, respectively. The explicit
expression of the angles-matrix for general asymptotic metric 
is
$$
A^{\alpha\bar\beta}=
{g_\infty^{i\gamma, j\bar\delta}p^\alpha{}_{i\gamma}
q^{\bar\beta}{}_{j\bar\delta}\over \sqrt{g_\infty^{i\epsilon,
j\bar\zeta}p^\alpha{}_{i\epsilon} p^{\bar\alpha}{}_{j\bar\zeta}}
\sqrt{g_\infty^{i\epsilon', j\bar\zeta'}q^\beta{}_{i\epsilon'}
q^{\bar\beta}{}_{j\bar\zeta'}}\ . }
\eqn\cyangles
$$

\section{Special Cases}

There are special cases of the geometries \ktgeom\ which admit
 two commuting KT structures.
To find these, we introduce a complex structure 
$$
\eqalign{
I(dz^1)&=dz^2
\cr
I(dz^2)&=-dz^1 
\cr
I(d\bar z^1)&=d\bar z^2
\cr
I(d\bar z^2)&=-d\bar z^1 }
\eqn\ancomstr
$$
on $\bC^2$
and the induced complex structure ${\bf I}$ on $\bC^{2k}$ as
in previous sections. The complex structure ${\bf J}$ commutes
with ${\bf I}$. We have seen that \ktgeom\ admits a KT structure
with respect to $(\nabla^{(+)}, {\bf J})$. For this geometry
 to admit another KT structure with
 respect to $(\nabla^{(-)}, {\bf I})$, the
asymptotic metric should be also hermitian with respect to ${\bf I}$
and 
$$
I\, d\tau=d\tau\, {\bf I}\ .
\eqn\ccseven
$$ 
The former condition is easily
met for a suitable choice of asymptotic metric. The 
latter condition
is satisfied provided that
$$
I\,p_i=p_i\,I\ ,
\eqn\cceight
$$
where 
the rotational parameters of the  maps $\tau$ have been
expressed in terms of complex 
$2\times 2$ matrices $\{p_i; i=1,\dots,k\}$. This
condition implies that
$$
p_i=\pmatrix{u_i& v_i\cr -v_i& u_i}
\eqn\paraparaa
$$
where $\{u_i, v_i; i=1,\dots, k\}$ 
are complex numbers. Therefore,
all the geometries \ktgeom\ associated with  maps that have
rotational parameters  \paraparaa\ 
admit two commuting KT structures. We remark that due to the
choice of the complex structure $I$, the dimension of these geometries
is always a multiple
of four.

\chapter{Supersymmetric Sigma Models}

Most of the  applications of the geometries
 that we have constructed
in the previous sections are in the context of two-dimensional
supersymmetric sigma models [\hull, \howea]. It is well
known that bosonic two-dimensional
 sigma models
with target space manifolds that have one KT structure admit an  
extension with off-shell (2,0) or (2,1) 
supersymmetry. If the target space
of such a sigma model admits another KT 
structure 
that commutes with the first 
, then the bosonic
sigma model admits an extension with off-shell (2,2) 
supersymmetry.
Next, bosonic sigma models with target 
space manifolds that have an HKT
structure admit an extension with off-shell
(4,0) or (4,1) supersymmetry. If  the  target space also admits
 a KT structure that commutes with the HKT structure
, then the sigma model admits an extension
with off-shell (4,2) supersymmetry. Finally, 
bosonic sigma models with target space
manifolds that have two commuting HKT structures 
admit a supersymmetric 
extension with off-shell (4,4) supersymmetry.
For example, sigma models with target spaces
the HKT geometries constructed in sections (2.1)-(2.3)
admit a (4,0)$\&$ (4,1), (4,2) or (4,4) supersymmetric extension
depending on whether the rotational parameters of the
 maps are quaternions, complex or real numbers,
respectively. Sigma models with target spaces
the KT geometries constructed in sections (5.1)-(5.3) admit
a (2,0)$\&$(2,1) or (2,2) supersymmetric extension depending
on whether the rotational parameters are generic or satisfy
\paraparaa, respectively.

All supersymmetric sigma models that we have described above admit an off-shell
constrained superfield formulation.  The construction
of these superfields as well as their actions can be easily done
following the results of [\howea]. 
Quantum mechanically, the off-shell (4,q)-supersymmetric
 two-dimensional sigma models are ultraviolet finite. Therefore
the eight-dimensional HKT geometries that we have constructed
 describe consistent string backgrounds. However, it is
expected that the sigma models
with (4,0) supersymmetry may receive $\alpha'$ corrections due
to the cancellation of the sigma model anomaly [\howeb]. The 
(2,q)-supersymmetric, $q\leq 2$, sigma models
have non-vanishing $\beta$-function even at one loop.
Related to this,  the eight-dimensional ($k=2$)
 KT geometries constructed in (5.1)-(5.3) do not seem to have a
brane interpretation. This is because they do not solve the 
IIA supergravity field equations which are the 
$\beta$-function vanishing conditions
at one loop. It would be of interest to find the K\" ahler-like
potentials [\hull, \hullwitten]  associated with 
the above KT and HKT geometries.

We summarize some of our results on the applications
of our geometries to sigma models 
in the table below. 
 \vskip 0.3cm
$$
\vbox{\settabs 4\columns
\+{\rm Supersymmetry} & \qquad $\nabla^{(+)}$ & 
$\nabla^{(-)}$ &{\rm Geometry} \cr
\+{$(4,4)$}&\qquad$Sp(k)$& $Sp(k)$& 
{\rm HKT} \& {\rm HKT}\cr
\+ {$(4,2)$}&\qquad $Sp(k)$& $SU(2k)$&{\rm HKT} \& {\rm KT}\cr
\+ $(4,0)\&\ (4,1)$ &\qquad$Sp(k)$& $SO(4k)$&{\rm HKT} \cr
\+ $(2,2)$&\qquad $U(2k)$& $U(2k)$&{\rm KT} \& {\rm KT}\cr
\+ $(2,0)\& (2,1)$ &\qquad$U(2k)$& $SO(4k)$&{\rm KT} \cr}
 $$
\vskip 0.2cm
\noindent{\bf Table 2: Sigma Model Geometries}\ \
This table contains (i) the supersymmetry
preserved by the 
two-dimensional sigma models, (ii) \& (iii) the holonomy
of the   $\nabla^{(+)}$ and $\nabla^{(-)}$ 
connections,  and (iv) the
type of geometry that the sigma model target space has
with respect to $\nabla^{(+)}$ and $\nabla^{(-)}$ , respectively.
\noindent

\chapter{Other superpositions}
\section{KT geometries and quaternions}

There are many other ways to superpose
the  four-dimensional HKT geometry \hktonea\
to construct new geometries in d-dimensions.
We can achieve this by appropriately choosing 
maps $\tau\, :\,
\bE^{d}\rightarrow \bE^4$. However, a generic choice of  maps $\tau$
will lead to  geometries which do not admit  KT or HKT structures
(see also the conclusions). However, it is possible to choose the
maps $\tau$ in such a way that the moduli space of the resulting 
geometries is  constructed using certain Grassmannians. To present an example
of such geometry, 
 we consider the  maps
$$
\tau:\quad \bH^k\rightarrow \bH
\eqn\ttnine
$$
 such that
$$
q\equiv \tau(u)=\sum_i p_i u^i  \bar c_i -a
\eqn\qlinearb
$$
where $(u^1,\dots,u^k)\in \bH^k$, $q\in \bH$, 
and $\{p_1,\dots,p_k; c_1, \dots, c_k; a\}$ are quaternions
that parameterize the  map. The geometry
that describes the superposition of 
four-dimensional HKT geometries in this
case is again obtained by pulling 
back \hktonea\ and summing over
the  maps $\tau$. We can also
 add $ds^2_\infty$ in the metric to
control its asymptotic behaviour.
For generic
choices of the parameters $\{p_1,\dots,p_k; c_1,
 \dots, c_k\}$, the
geometry that describes the superposition
 will not admit any KT structures.

${(i)}:\,$ A special case of the above geometry is to allow either
the rotational parameters $\{p_1,\dots,p_k\}$ or the
rotational parameters $\{c_1,
 \dots, c_k\}$ of the maps $\tau$ to be complex numbers.
Let us suppose that $\{ c_i=c_i^0+i c_i^1; i=1, \dots, k\}$;
the other case is symmetric. Then the geometry which
is associated with this superposition will admit
a KT structure with respect to $(\nabla^{(+)}, {\bf J}_1)$.
Two-dimensional sigma
models with bosonic couplings
 determined by this geometry 
 admit a (2,0)
 off-shell supersymmetric extension.

${(ii)}:\,$ Next,  let us take all the rotational parameters
$\{p_1,\dots,p_k; c_1, \dots, c_k\}$ of
the  maps $\tau$ to be complex numbers, i.e.
 $\{p_i=p_i^0+ip_i^1; i=1,\dots,k\}$ 
 and $\{ c_i=c_i^0+i c_i^1; i=1, \dots, k\}$.
 We remark that this construction  can be done with
other combinations of complex structures. 
In this case, 
we can show that the geometry that describes
the superposition admits two KT structures, 
$(\nabla^{(+)}, {\bf J}_1)$ and   
$(\nabla^{(-)}, {\bf I}_1)$.
This geometry is related to that given in section (5.3)
but the choice of complex structures is different. 
Since the complex structures
${\bf I}_1$ and ${\bf J}_1$ commute, 
 two-dimensional sigma
models with bosonic couplings
 determined by this geometry 
admit a (2,2)
 off-shell supersymmetric extension.

\section {Moduli and Calibrations}

Maps $\tau$ and $\hat \tau$ that are equivalent under the
relation
$$
\tau\sim \hat\tau\ ,
\eqn\cvone
$$
where  
$$
\{\hat p_1,\dots,\hat p_k; \hat c_1,
 \dots,\hat c_k; \hat a\}=\{s p_1 ,\dots,s p_k; v c_1,
 \dots,v c_k; s a \bar v\}\ ,
\eqn\cvtwo
$$
$s,v\in \bH$, lead to the same geometries. The space of equivalence
classes of such maps is the bundle space $E(\theta_1^{k-1})$ of a
 quaternionic
line bundle over 
$$
Gr(1, \bH^{k})\times Gr(1, \bH^{k})\ .
\eqn\cvthree
$$
The quaternionic line bundle $\theta_1^{k-1}$  restricted to the
first  quaternionic projective space is isomorphic to the associated
canonical quaternionic line bundle, and $\theta_1^{k}$ restricted
to the second quaternionic projective space is isomorphic to the
associated conjugate canonical quaternionic line bundle.
The moduli space of the above geometries associated with N distinct
linear maps $\tau$ is
$$
{\cal M}_N={D_N\times\big(\times^N E(\theta_1^{k-1})-\Delta\big)\over S_N}\ .
\eqn\cvfour
$$

${(i)}:\,$ In the case above for which half of the rotational parameters of
the maps $\tau$ are complex, the equivalence classes of maps is the
bundle space $E(\kappa)$ of a quaternionic line bundle $\kappa$ over
$$
Gr(1, \bH^{k})\times \bC P^{k-1}\ .
\eqn\cvfive
$$
The moduli space of the above geometries associated with N distinct
linear maps $\tau$ is
$$
{\cal M}_N={D_N\times\big(\times^N E(\kappa)-\Delta\big)\over S_N}\ .
\eqn\cvs
$$
It is clear that this geometry is associated with the K\"ahler calibration
constructed out of ${\bf J}_1$.

${(ii)}:\,$ Finally, in the case for which all the rotational  parameters of $\tau$
are complex numbers, the equivalence classes of maps is the bundle space
$E(\lambda)$ of a
 quaternionic line bundle $\kappa$
 over
$$
\bC P^{k-1}\times  \bC P^{k-1}\ .
\eqn\cpcp
$$
The bundle $\lambda$ can be constructed using 
the  canonical complex line bundle of $\bC P^{k-1}$. 
The moduli space of the above geometries associated with N distinct
linear maps $\tau$ is
$$
{\cal M}_N={D_N\times\big(\times^N E(\lambda)-\Delta\big)\over S_N}\ .
\eqn\cvseven
$$

The eight-dimensional ( $k=2$) KT geometry above   is also
 associated with one of the calibrated
geometries of [\calic].  To see this, we write the kernel of the
maps $\tau $ as
$$
\sum_{i=1}^2{p}_iu^i\bar c_i - a = 0 \ .
\eqn\ktrl
$$
We observe that the tangent spaces of the kernels are stabilized 
both by the action of ${\bf
I}_1$ and
${\bf J}_1$, so they are complex relative to these holomorphic structures.
Thus they are simultaneously calibrated by the K\"ahler 4-forms
associated with ${\bf I}_1$ and ${\bf J}_1$, and so also by 
$$
\tilde \Theta={1\over 4}(\omega_{{\bf I}_1}^2- \omega_{{\bf J}_1}^2).
\eqn\cveight
$$
The contact set of this calibration is 
$$
G_{\Lambda}={SU(2)\over S\big (U(1)\times U(1)\big)}\times {SU(2)\over
S\big(U(1)\times U(1)\big)}=S^2\times S^2\ .
\eqn\cvnine
$$
Note that again  the moduli space of these KT geometries is
constructed out of the contact set of the calibration.

\chapter{Conclusions}

We have presented new
HKT and KT
geometries by superposing the four-dimensional
HKT geometry associated with
the NS-5-brane. The  4k-dimensional HKT geometries
have been constructed  using superpositions along quaternionic 
planes in $\bH^k$. We have found that the moduli space
of these HKT geometries is constructed from the
 canonical quaternionic line
bundle over the  projective space $Gr(1,\bH^k)$.
Several special cases of these geometries were also
considered and their moduli spaces have been given. 
In addition, we have shown that our
eight-dimensional HKT geometries are superpositions
 along calibrated quaternionic planes in $\bH^2$. Their moduli
spaces can be  constructed using a 
quaternionic line bundle over the contact set
of the associated calibration.  All these 
eight-dimensional  geometries are  solutions
 of supergravity theories
with the interpretation of intersecting branes on 
a string preserving $3/32$, $1/8$, $5/32$ and
$3/16$ of spacetime supersymmetry. The proportion of 
supersymmetry preserved
is directly related to the class of calibrated planes used
to construct the superposition.  The $4k$-dimensional 
HKT geometries  have also
applications in the context of two-dimensional 
sigma models. In particular we have  
found new sigma models   with (4,0), 
(4,1) and (4,2) supersymmetry 
generalizing the  (4,4)-supersymmetric
sigma models of [\kelly].
We have also constructed new  
KT geometries using superpositions
of the same four-dimensional  
HKT geometry  
along complex planes
in $\bC^{2k}$.  The
resulting  KT geometries are associated with , but not 
uniquely determined
by, arrangements of complex planes in $\bC^{2k}$. In
 addition we investigated several other aspects
of these geometries leading, for example, 
to the construction of geometries associated
 with two-dimensional
sigma models with (2,0) and (2,2) supersymmetry. 
Several special cases have also 
been examined. In particular, we have found
that some of the
eight-dimensional KT geometries are also 
associated with calibrated
planes in $\bC^4$.

The investigation of the superposition of 
four-dimensional HKT geometries
 may be  extended further in two ways. First, 
there may be calibrations
in $4k$-dimensions, $k>2$, for which the planes,
that are used in the construction of the $4k$-dimensional,$k>2$,
HKT geometries, are calibrated.
These calibrations
will  generalize
those of [\calic] and some of the results we have found 
for the eight-dimensional
HKT geometries. 
Alternatively, we may be able to find new
 superpositions of four-dimensional HKT
geometries by choosing planes 
in $\bE^{4k}$ which are associated
with other calibrations like for example the
 Special Lagrangian calibration or the 
calibrations associated with the 
exceptional groups $G_2$ and ${\rm Spin}(7)$ [\calia, \calib].
It would be of interest to study 
the properties of the  geometries that result from
such superpositions and see whether they admit a brane
interpretation. 

In eight dimensions, it is clear that there is a  correspondence
between the type of calibration used to construct the HKT
geometry and the holonomies of the $\nabla^{(+)}$ and $\nabla^{(-)}$
connections. In particular the holonomy of $\nabla^{(+)}$ is
in all cases $Sp(2)$. This $Sp(2)$ fixes (up to the trivial
$Sp(1)$-factor) a quaternionic calibrating four-form. Adding to
this quaternionic four-form  a certain $Spin(7)$, $SU(4)$ or $Sp(2)$
invariant calibrating four-form results in $\nabla^{(-)}$ having 
$Spin(7)$, $SU(4)$ or $Sp(2)$ holonomy, respectively.
The understanding of this correspondence is a key for 
studying the geometric properties of the
superposed geometry and it may lead to many more 
applications in string and M-theory.

\vskip 1cm
\noindent{\bf Acknowledgments:}   G.P. is 
supported by a University Research Fellowship
from the Royal Society. 

\endpage

\appendix
\chapter{Symmetry, Supersymmetry and Five-branes}

The HKT geometry on $\bH-\{0\}$ associated
with the NS-5-brane can be rewritten as
$$
\eqalign{
ds_{(4)}^2&=\big(1+{ \mu\over |q|^2}\big) |dq|^2
\cr
H&={\mu\over 3!}{\rm Re}\big( {d\bar q\wedge dq\wedge (d\bar q q -
\bar q dq)\over 2
|q|^4}\big)\ ,}
\eqn\hktone
$$
where $q\in \bH$, $\bar q$ is the 
conjugate of $q$ and $|q|^2=\bar q q$. 
We can introduce a pair of hyper-complex structures
using left and right  multiplication by the imaginary
unit quaternions on $dq$ as in  \hheight\ and \rcom. Since the left 
and right actions
commute,  the triplets of complex structures also
commute leading to a commuting pair of HKT 
structures. The manifold  $\bH-\{0\}$ with \hktone\ 
can be the target space of a 
two-dimensional sigma model with (4,4) supersymmetry.

The metric and torsion of the
 HKT geometry \hktone\ are invariant under the
$SO(4)=Sp(1)\times Sp(1)/\bZ_2$ action
$$
q\rightarrow a q \bar b\ ,
\eqn\ggone
$$
where $(a,b)$ are
 quaternions and  $(a,b)$$\in$$Sp(1)\times Sp(1)$. 
This group action
does not commute with the  
maps associated with the complex structures, i.e. the
above isometries are not holomorphic.

Using this group action, we can 
construct new HKT geometries from that of \hktone. 
This can be achieved by factoring $\bH-\{0\}$
 with a discrete subgroup, $G$, 
of $Sp(1)\times Sp(1)$. Similar  orbifold-like
constructions have been done for D-branes in [\kach]. 
There are many choices
for
the subgroup $G$. However  we shall
 require the resulting manifold to admit at least one
HKT structure; there are other choices of $G$ that 
lead to manifolds with
one or two K\"ahler structures with torsion.
 This restricts
the choice of
$G$ to be a subgroup either of the 
$Sp(1)$ that acts from the left or the $Sp(1)$
that acts from the right. If we 
choose $G$ to be a subgroup of $Sp(1)$ that acts
from the left, then the manifold
 $\bH-\{0\}/G$ will admit an HKT structure
with respect to the pair $(\nabla^{(+)}, J_r)$. 
We remark that
$G$ acts on $\bH-\{0\}$ freely since
 the only fixed point of the left $Sp(1)$
action on $\bH$ is  $q=0$ but this
 point is not part of the space
$\bH-\{0\}$. The manifold  $\bH-\{0\}/G$  
can be the target space of a 
two-dimensional sigma model with (4,0) or (4,1)
off-shell supersymmetry.

 A special case of the above 
construction arises 
whenever the discrete subgroup, 
$G$, of $Sp(1)$ is chosen such that
 it preserves one of
the
$I_1, I_2$ or $I_3$ complex 
structures. If $G$ preserves say
the $I_1$ complex structure, then 
the action of $G$ commutes with the  map
induced by the $I_1$ complex structure. This 
requires that the elements $a\in G\subset
Sp(1)$ are of the form
$$
a=a_0+i a_1\ .
\eqn\ggtwo
$$
Next since $\bar a a=1$, then 
$$
a=\cos\theta+i \sin\theta\ ,
\eqn\ggfour
$$
where $\theta$ is an
angle. For example,  we choose $G=\bZ_N$ 
acting on $\bH-\{0\}$ with $e^{
{2\pi k i\over N}}$. The manifold
 $\bH-\{0\}/G$ admits an HKT structure with respect to
the pair $(\nabla^{(+)}, J_r)$. 
In addition it admits a K\"ahler structure
with
torsion with respect to the complex 
structure $I_1$.  The manifold  $\bH-\{0\}/G$  
can be the target space of a 
two-dimensional sigma model with (4,2) 
off-shell supersymmetry.

\refout

\bye